# Out-of-Plane Mechanical Properties of 2D Hybrid Organic-Inorganic Perovskites by Nanoindentation

Qing Tu[1,2], Ioannis Spanopoulos[3], Shiqiang Hao[1], Chris Wolverton[1], Mercouri G. Kanatzidis[3], Gajendra S. Shekhawat[2*], Vinayak P. Dravid[1,2*]

*1. Department of Materials Science & Engineering, Northwestern University, Evanston, IL 60208, USA*

*2. Northwestern University Atomic and Nanoscale Characterization Experimental (NUANCE) Center, Northwestern University, Evanston, IL 60208, USA*

*3. Department of Chemistry, Northwestern University, Evanston, IL 60201, USA*

**Abstract**

2D layered hybrid organic-inorganic perovskites (HOIPs) have demonstrated improved stability and promising photovoltaic performance. The mechanical properties of such functional materials are both fundamentally and practically important to achieve both high performance and mechanical stable (flexible) devices. Here we report the mechanical properties of a series of 2D layered lead iodide HOIPs and investigate the role of structural sub-units (*e.g.,* variation of the length of the organic spacer molecules -R and the number of inorganic layer -n) on the mechanical properties. While 2D HOIPs have much lower nominal elastic moduli and hardness than 3D HOIPs, larger n number and shorter R lead to stiffer materials. DFT simulations showed a similar trend to the experimental results. We compared these findings with other 2D layered crystals and shed light on routes to further tune the out-of-plane mechanical properties of 2D layered HOIPs.

**Introduction**

Hybrid organic-inorganic perovskites (HOIPs), most notably three-dimensional (3D) $AMX_3$ (A = small organic cation , M = divalent group 14 element and X = halogen) materials, have attracted remarkable research interest in the photovoltaic community in recent years because of their high solar power conversion efficiency (now >22% in laboratory conditions) and the ease of solution-based processing method.[1-2] Most of the research has been devoted to understand the mechanism of exciton generation and transport to further improve the photovoltaic efficiency of solar cells based on these materials by controlling the chemical composition, crystal structure, and design of the corresponding materials and devices.[3-5] However, 3D HOIPs have suffered from low stability under ambient environmental conditions in the presence of moisture, oxygen, and UV light, thus limiting their application in commercial photovoltaic devices.[6-7]

Two-dimensional (2D) layered HOIPs are the lower dimensional structural derivatives of 3D HOIPs, whose structure can be derived from the ordered removal of the M-component from the inorganic framework along certain direction in the parent 3D perovskite structure and replacing the removed inorganic layers by organic layers.[8] The resulting 2D HOIPs features a structure with alternating organic and inorganic layers, where the neighboring organic layers interact with each other through weak Van der Waals interactions. Figure 1 shows an example of 2D layered HOIPs, which is currently the richest subgroup of the 2D perovskite family.[8] These particular compounds belong to the so-called Ruddlesden−Popper family of materials,[9-11] exhibiting a general formula of (RNH$_3$)$_2$MA$_{n-1}$Pb$_n$I$_{3n+1}$, where R is the organic spacer molecule, MA is the methylammonium cation, and n indicates the number of inorganic layers, i.e., the thickness of the inorganic part of the framework. Because of this unique structure, 2D HOIPs have been demonstrated with improved stability and promising photovoltaic performance.[12-14] Furthermore, this 2D perovskite structure also relaxes the steric constraints on the inorganic cations, as outlined by the Goldschmidt's tolerance factor in 3D HOIPs,[15] providing a much larger compositional space to engineer new materials with tailored properties.[8, 16-18]

The research on the physical properties and electronic structures of HOIPs has made rapid progress.[5, 19] The scalable solution-based processing methods of high-quality HOIP thin film devices allow their potential applications to go beyond photovoltaics into flexible electronics.[20-21] The mechanical stress state of HOIPs in these applications can significantly impact their electronic properties and mechanical robustness of HOIPs-based devices.[22-24] For 3D HOIPs, there have been some theoretical and experimental studies of the mechanical properties of the materials. Feng et al.[25] first reported the elastic modulus of CH$_3$NH$_3$MX$_3$ (M = Sn, Pb, and X = Br, I) based on density functional theory (DFT) calculations. Cheetham and coworkers measured the static Young's modulus and hardness of CH$_3$NH$_3$PbX$_3$ (X = Cl, Br, and I) crystals by nanoindentation.[26] They found a clear correlation of the Young's modulus to the Pb-X bond strength. Similar conclusions were reached in the nanoindentation study by Cahen et al.,[27] where they found changing the A ion from organic CH$_3$NH$_3^+$ to inorganic Cs$^+$ has negligible effects on Young's modulus while the Pb-X strength is the dominating effect. Reyes-Martinez et al.[28] further showed that the dynamic mechanical properties of APbX$_3$ (A = Cs, CH$_3$NH$_3$; X = I, Br) and found that the organic molecules in HOIPs surprisingly do not affect the time-dependent mechanical properties either. Other 3D HOIP single crystals with different chemical composition and along different crystal orientations have been investigated by nanoindentation and ultrasonic methods.[29-32] The measured Young's moduli are usually about 10 to 30 GPa, which is significantly more compliant than other inorganic perovskite materials like BaTiO$_3$[33] and Pb[Zr$_x$Ti$_{1-x}$]O$_3$.[34] It is believed that this compliant nature of HOIPs results in low energy barrier to structural change, such as phase transition, ion migration and rearrangement of point defects ("self-healing"),[3, 5, 8, 20] which is critical to understand the physical process in photovoltaic applications. Being compliant also provides manufacturing convenience to perovskite thin film formation on surfaces with different morphologies by scalable methods.[26-27]

However, despite the importance of understanding the mechanical behavior, there are hitherto no reports on the mechanical properties on 2D HOIPs, especially how the dimensional reduction from 3D to 2D layered structure and the structural parameters (i.e., nature of organic spacer, R-NH$_2$, and number of inorganic layers, n) affects the resulting mechanical properties of the 2D layered HOIPs. Those compounds are among the latest addition to the 2D material family,[35] which are ionically or covalently bonded in-plane while out-of-plane bonded by weak Van der Waals forces. Like in many other 2D layered crystals such as highly ordered pyrolytic graphite (HOPG), hexagonal boron nitride (h-BN) and molybdenum disulfide (MoS$_2$), the in-plane mechanical property is intrinsic to the in-plane bond strength,[36] the out-of-plane mechanical property (*i.e.,* perpendicular to the layered material planes) is subjected to changes in interfacial interaction between layers,[37-39] and thus is more sensitive to this structural and interfacial organic molecular changes. Here we report the study of the static mechanical properties of a series of 2D Ruddlesden−Popper HOIPs in the out-of-plane direction by nanoindentation. To exclude the effect of other factors such as M-X bond strength and complex interactions between R functional groups, we focus our study on lead iodide perovskite using only linear aliphatic organic spacer molecules (R-NH$_2$). DFT simulations are used to provide theoretical insights into the observed trend. We discuss the obtained results in the broad context of 2D layered crystals and shed light on methods to further engineer the mechanical properties of 2D layered HOIPs.

**Materials and Methods**

*Materials*

We synthesized 7 different types of 2D HOIPs materials in single crystal form, exhibiting various number of inorganic layers n and different organic spacer molecules R-NH$_2$, with a general formula of $(C_mH_{2m+1}NH_3)_2(CH_3NH_3)_{n-1}Pb_nI_{3n+1}$. For simplicity purposes the following sample notation will be used throughout the manuscript, *e.g.*, C4n1. The number after C indicates the number of carbon atoms in the organic spacer molecule chain, and the number right after n indicates how many inorganic layers (composed of corner-shared PbI$_6$ octahedra) lies between two adjacent organic layers. For instance, *C4n1* suggests the sample has R = -C$_4$H$_9$, and n = 1. To investigate the influence of n number on the out-of-plane mechanical properties, R was kept constant (here R = -C$_4$H$_9$) and we varied the n = 1, 3 and 5 (Figure 1). When studying the effects of the length of R, we kept n = 1 and varied the R = C4, C5, C6, C8, C12 (Figure 2). Details on the synthesis of the aforementioned compounds can be found in Supporting Information (SI) – Section

I. All samples were characterized by powder X-ray diffraction measurements (PXRD) and the resulting patterns were compared with the calculated ones form the solved single crystal structures (see SI – Section II, Figs. S1 to S7) to verify the phase purity. Extensive analysis of the single crystal structures of all the examined 2D perovskites are reported elsewhere.[40-44] Large single crystals (lateral size ~ a few mm and thickness ranging from 100 µm to about 1 mm) are acquired for nanoindentation experiments.

*Nanoindentation*

Hardness and indentation modulus were measured using a Hysitron 950 Triboindenter with a Berkovich indenter (three-sided pyramid shape diamond tip, tip radius ~ 100 nm) in ambient environment. Large single crystal flakes of 2D HOIPs were first mounted onto stainless steel atomic force microscopy specimen discs by epoxy (J-B Kwik) with the (001) plane facing up. Prior to nanoindentation, the top few layers were mechanically exfoliated by scotch tape to remove any potential surface contamination/degradation and reveal fresh crystal plane.

For all measurements, the loading and unloading were kept at about 70 µN/s and before unloading the indenter was hold at constant load for 30s. The peak load was kept at 1000 µN for samples with different n number while for samples with different spacer molecules, the peak load was kept at 500 µN so that the indentation depth can be kept within 500 nm. On C4n1 and C5n1 samples, we have tested with both 500 µN and 1000 µN peak loads and did not see a difference in the measured mechanical properties. The data were analyzed using standard Oliver and Pharr analysis to extract reduced moduli and hardness.[45-46] The out-of-plane Young's moduli $E_\perp$ of the materials can be further derived with a Young's modulus *E* of 1141 GPa and Poisson's Ratio $\nu$ of 0.07 for diamond tip[46] and $\nu$ = 0.3 for HOIPs.[26-27] More than 3 crystals were checked for each type of sample and 10 to 20 indentations were performed on each crystal and the reported values for each type of samples were averaged by these measurements.

**Results and Discussion**

The stiffness measured by nanoindentation is strongly dominated by the elastic response along the indenter axis and is only weakly affected transversely.[46-47] Thus nanoindentation can provide a reliable measurement of the out-of-plane mechanical property of the 2D HOIPs. The typical load-indentation curves (*P-h*) curves are shown in Figure 3 for each type of sample. Some small discontinuities in the loading curves, known as "pop-in" events, are present. Similar pop-in events are also reported in nanoindentation on 3D methylammonium, cesium and formamidinium lead halide perovskite single crystals,[28, 31] and are very common in nanoindentation study of other 2D layered crystals (*e.g.,* HOPG[48] and muscovite mica[49]). The abrupt change in indentation depth in pop-in events is attributed to plastic deformation caused by nucleation and propagation of slip dislocations in 3D system,[28, 31] while in 2D crystals, it can also be caused by layer kinking, layer delamination and layer fracture.[48, 50] We find that the pop-in events are more common in samples with low n number, which indicates easier slippage between Van der Waals bonded layers and/or aforementioned layer deformations as reducing the relative amount of inorganic compositions in the crystals. The observed plastic deformation suggests good ductility of these 2D layered HOIPs, which can benefit their applications in flexible electronics to prevent catastrophic fracture/disintegration during deformation.[51] How the plastic deformation will affect the electronic properties of 2D and 3D HOIPs is essential to these flexible electronic applications[20-21] and strain engineering of these materials,[25, 51] which calls for investigations in the future.

Analysis of the unloading *P-h* curves by standard Oliver and Pharr method can extract the Young's modulus and hardness of the samples. Figure 4 clearly shows the dependence of the out-of-plane mechanical property on the crystal structures. While keeping R = -$C_4H_9$, both the Young's moduli and hardness increase as n increases from 1 to 5. C4n1 samples have the lowest Young's modulus (3.3 ± 0.1 GPa), which is about 1/3 to 1/5 of the Young's modulus *E* of tetragonal $CH_3NH_3PbI_3$ single crystal along <100> direction reported in literature.[26-27, 32] The hardness of C4n1 samples is also much smaller than that of 3D crystal (about 30% to 50%).[26-27, 32]

In 3D organic-inorganic halide perovskite single crystals, the Young's modulus and hardness are determined by the ionic *M-X* (*M = Pb, Sn*; *X = Cl, Br*, or *I*) bond strength.[25-27] In 2D perovskite, some of the inorganic layers are replaced by soft, organic layers (here alkyl layers) compared to 3D perovskites. Furthermore, these organic layers interact with each other through Van der Waals forces, which are orders of magnitude weaker than ionic bonds. Our results show that in 2D structure, the presence of organic layers with weak Van der Waals interfaces in the crystals, can significantly soften HOIPs, which can provide a route to engineer more compliant HOIP materials to facilitate their applications in flexible devices. As n increases from 1 to 5, the softening effect from the organic layers in 2D HOIPs decreases, and thus the values of Young's modulus and hardness increase and approach the values measured from 3D $CH_3NH_3PbI_3$ crystals. Similar trends in the dependence of Young's modulus on n are captured by first principle DFT simulations, as shown in SI – Section III.

We further investigated the effects of the organic molecule spacer R on the mechanical properties of 2D HOIPs (Figures 3 and 5). Here we kept n = 1, and changed the alkyl chain length in R, starting from -$C_4H_9$, to -$C_{12}H_{25}$ (Figure 2). Intuitively, as the alkyl chain length increases and hence the relative amount of organic part (softer than the stiff inorganic materials) increases, one would expect a decrease in Young's moduli of the materials. In fact, as R increases from C4 to C6, the out-of-plane Young's modulus of 2D HOIPs gradually decreases. In contrast, organic cations in 3D HOIPs have minimum effects on either static or dynamic mechanical properties, as evidenced by the fact that changing the A site ion from inorganic $Cs^+$ to organic $CH_3NH_3^+$ results in negligible changes in mechanical properties[26, 28] and the metal-halide bond strength determines the mechanical characteristics of these materials.[25-28]

However, this softening effect due to the increasing alkyl chain length saturates eventually (Fig. 5A), which suggests the $PbI_6^{4-}$ inorganic layer still plays a significant role in determining the out-of-plane mechanical properties of 2D HOIPs. This is even more clear in hardness, where we did not find a clear dependence on the spacer molecules (Fig. 5B). Hardness measures the resistance of the materials to plastic deformation and inorganic materials usually have higher resistance to plastic deformation than organic materials. For time-dependent properties such as viscoelasticity and stress relaxation, the organic layer, especially for 2D HOIPs with long chain spacer molecules like -$C_{12}H_{25}$ and -$C_{18}H_{37}$,[52] should play a more important role than inorganic layer 2D HOIPs, which is beyond the scope of our study.

The out-of-plane Young's modulus $E_\perp$ of many 2D, Van-der-Waals bonded crystals have been experimentally investigated. Table I summarizes the literature reported $E_\perp$ values for some most widely studied 2D crystals, including HOPG,[53-54] muscovite mica,[55-57] $MoS_2$,[58] h-BN,[59-60] and $Ti_3C_2T_x$ (MXene).[38] Compared to these 2D crystals, 2D HOIPs investigated here have the lowest elastic moduli in the out-of-plane direction. This compliant feature of HOIPs gives more tunability of structure and properties by external mechanical, thermal and electrostatic stimuli. Changing HOIPs from 3D metal-halide framework to 2D Van-der-Waals bonded crystals further significantly softens the materials. Similar changes have also been found in 3D $Ti_3SiC_2$ crystal to 2D MXene phase. After dissolving the Si atoms, replacing the strong covalent bonding by weak Van der Waals and hydrogen bonding between layers, $E_\perp$ decreases from 320 GPa[61] to about 12 ~ 70 GPa.[38] $E_\perp$ can be further tuned through the interfaces between each 2D layers. As shown in 2D MXene, water intercalation into the layers further decreased $E_\perp$ to 2 ~ 12 GPa.[38] For graphene/graphite, experimental study has reported that reducing the interfacial interaction strength can reduce its stiffness in out-of-plane direction[37] while molecular dynamics simulation predicted that intercalation of graphite by stiff $Li_xC$ ions can increase $E_\perp$ by 4 times.[62] Thus $E_\perp$ of 2D HOIPs can be further engineered by choosing R with different functional groups and/or structures to have weaker or stronger layer-layer interactions.

**Summary & Conclusions**

We report out-of-plane mechanical properties of a series of 2D hybrid organic-inorganic lead iodide crystals as a function of the number of inorganic layers n and the alkyl chain length of the organic space molecule R using nanoindentation. We find the 2D HOIPs are softer than their 3D counterparts due to the replacement of the ionic bonds by the soft organic layers and the weak Van der Waals interactions. As n increases from 1 to 5, the relative amount of these weak factors in the crystals are decreasing and both the Young's modulus and hardness increase, approaching to the reported values of corresponding 3D crystals. Furthermore, we show that increasing the alkyl chain length from C4 to C12, $E_\perp$ first decreases and eventually plateaus while no clear trend in hardness is found. These results demonstrate that variations in the sub-units of these 2D HOIPs, like n and R, have a large effect on their mechanical properties. The competition between the stiff inorganic layers, the soft organic layer and the weak Van der Waals interface determines the mechanical properties of those materials.

Compared to the other 2D Van der Waals bonded crystals, 2D HOIPs are among the most compliant of them all. Insights from such mechanical property studies of these crystals, especially along the out-of-plane direction, suggest that one can further tune the mechanical properties of 2D HOIPs through interface engineering. The mechanical behavior of 2D HOIPs is an important design parameter towards stable solar cell systems, manufactural processes and other specific applications like flexible electronics/photovoltaics. The study of the out-of-plane mechanical property can especially facilitate the understanding of interfacial interaction between 2D layers and can correlate with the out-of-plane carrier and phonon transport, thus providing valuable insights into the structure-property relations in this type of materials. These findings will allow the elaborate design of devices with desired mechanical and electronic properties.

**Supporting Information**

Details about synthesis, XRD characterization and DFT simulations.

**Author Information**

*Corresponding authors:

Dr. Gajendra S. Shekhawat: g-shekhawat@northwestern.edu

Prof. Vinayak P. Dravid: v-dravid@northwestern.edu

**Acknowledgement**

This work made use of the SPID facilities of Northwestern University's NUANCE Center, which has received support from the Soft and Hybrid Nanotechnology Experimental (SHyNE) resource (NSF ECCS-1542205); the MRSEC program (NSF DMR-1720139) at the Materials Research Center; the International Institute for Nanotechnology (IIN); the Keck Foundation, and the State of Illinois, through the IIN. This work was supported by the National Science Foundation IDBR Grant Award Number 1256188, and partially supported by Air Force Research Laboratory grant FA8650-15-2-5518. M.G.K. acknowledge the support under ONR Grant N00014-17-1-2231.

# Figures

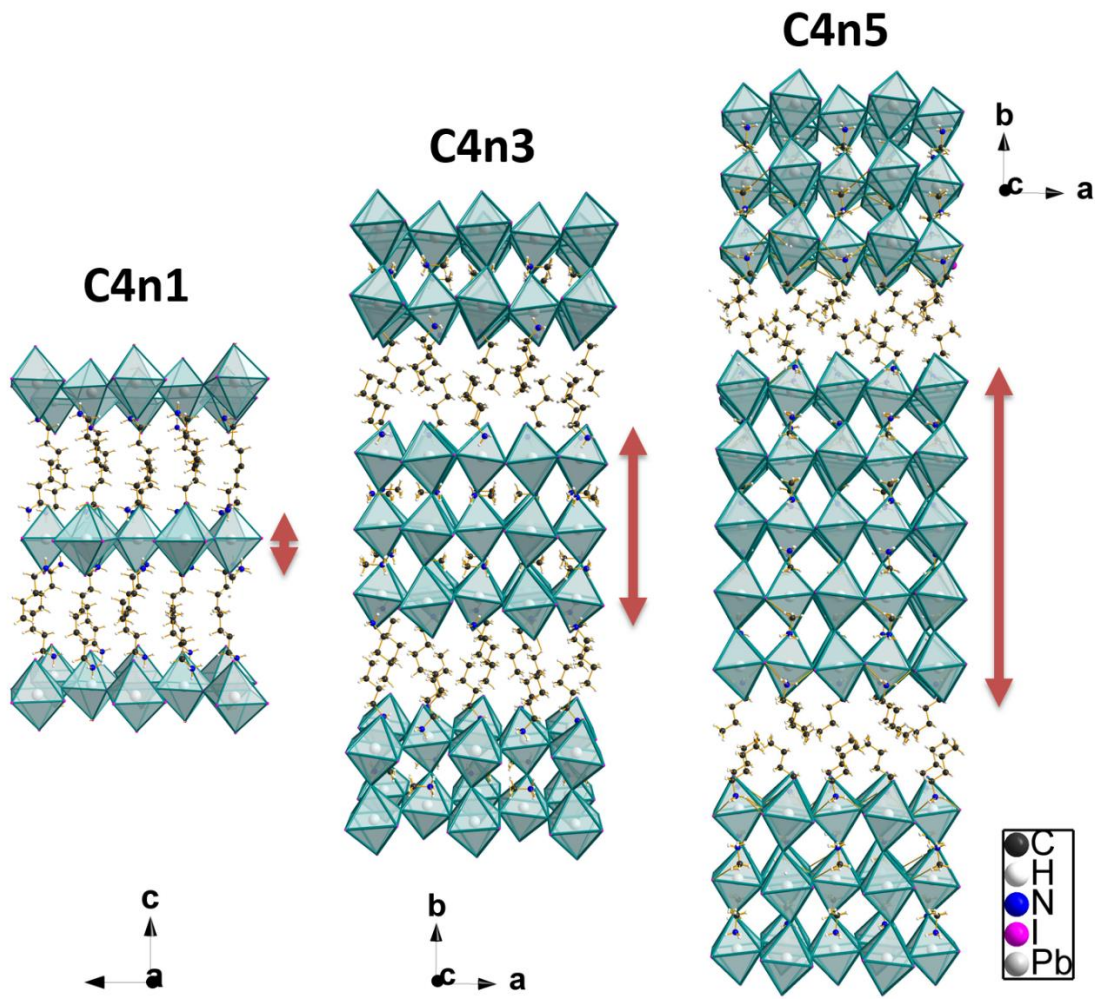

**Figure 1.** Part of the 2D structures of the layered $(CH_3-CH_2-CH_2-CH_2-NH_3)_2(CH_3-NH_3)_{n-1}Pb_nI_{3n+1}$ family of materials with increasing number of inorganic layers from (n = 1, C4n1) to (n = 5, C4n5). The turquoise octahedra represent the $[PbI_6]^{4-}$ moieties. In the case of C4n3 and C4n5 methylammonium cations $(CH_3-NH_3^+)$ are residing among the inorganic layers, to charge balance the framework. The red arrow indicates the increase in the size of the inorganic part, that is the number of inorganic layers, ranging from 6.4 Å to 31.4 Å.

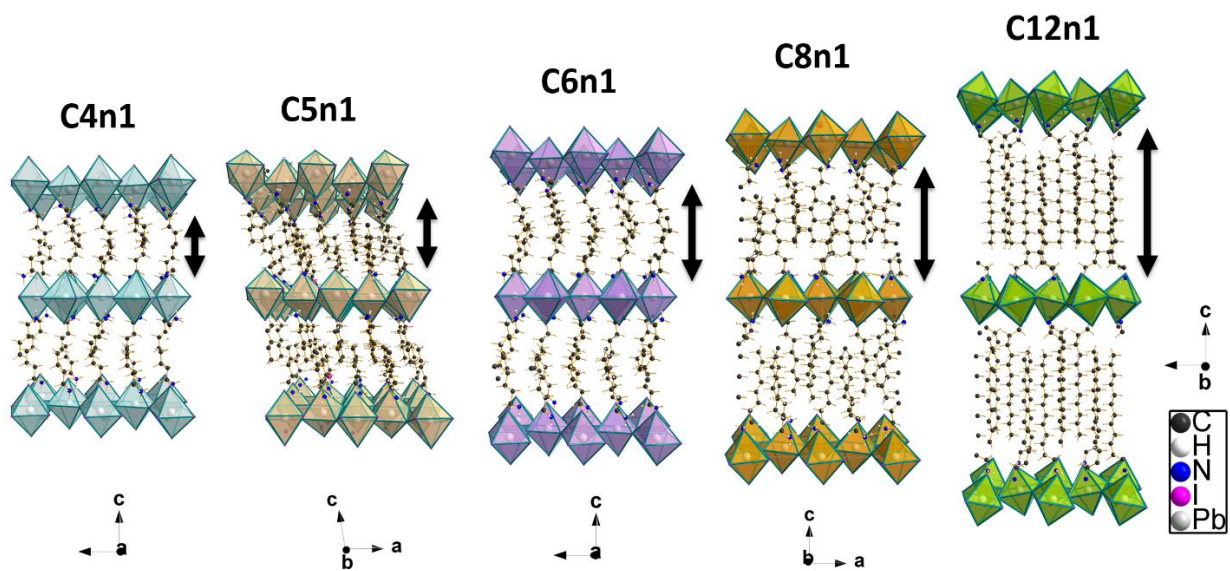

**Figure 2.** Part of the 2D structures of the layered (R-NH$_3$)$_2$PbI$_4$ materials with increasing length of the R-NH$_2$ organic spacer, (R: CH$_3$-(CH$_2$)$_3$- = C4, (CH$_3$-(CH$_2$)$_4$- = C5, (CH$_3$-(CH$_2$)$_5$- = C6, (CH$_3$-(CH$_2$)$_7$- = C8, (CH$_3$-(CH$_2$)$_{11}$- = C12, for the same number of inorganic layers, n = 1. The colored octahedra represent the [PbI$_6$]$^{4-}$ moieties. The black arrow indicates the increase in the distance of the inorganic layers with increasing the length of the organic spacer, ranging from 7.4 Å to 17 Å.

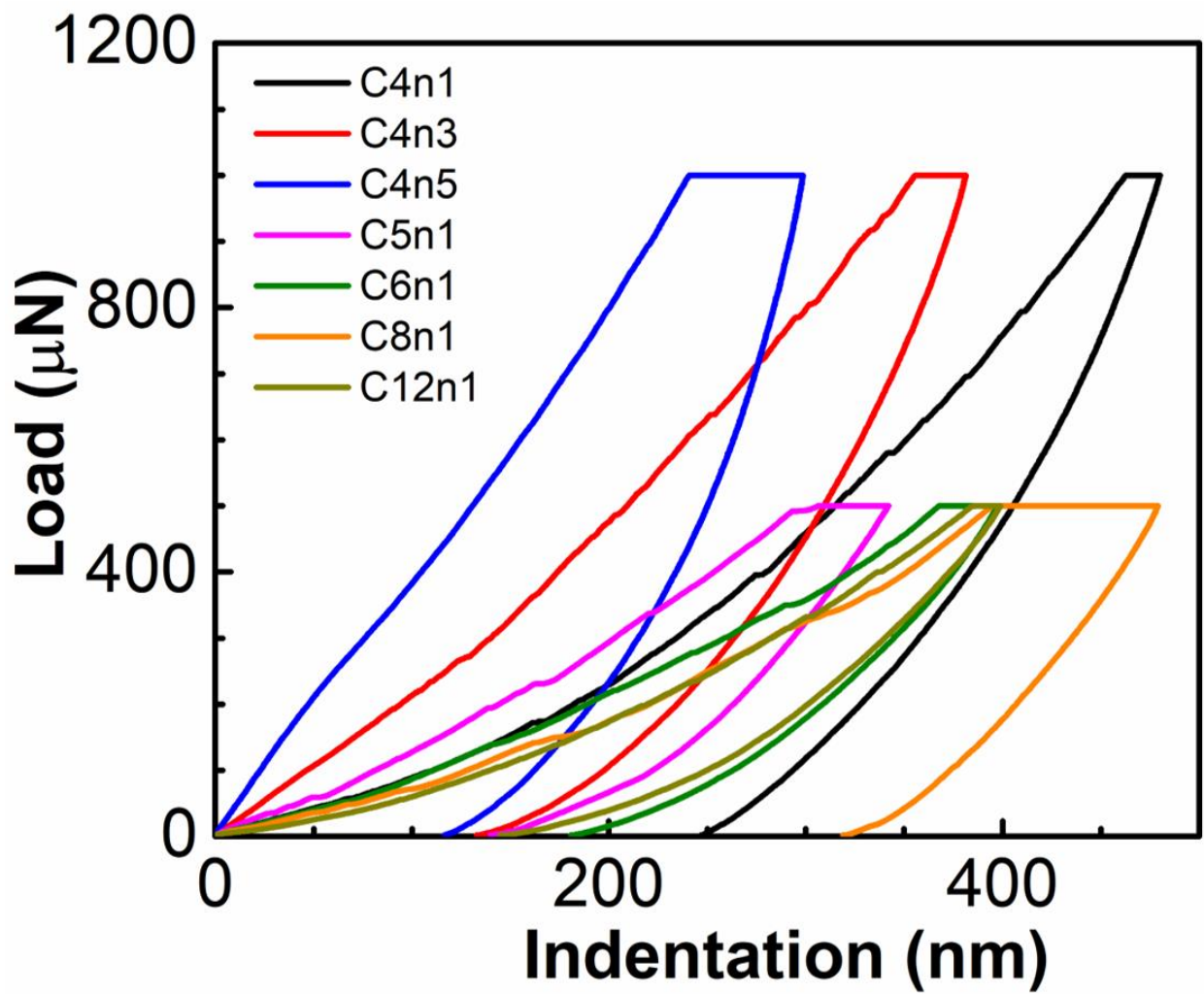

**Figure 3.** Typical load vs indentation (*P-h*) curves of each 2D layered HOIPs investigated in this study.

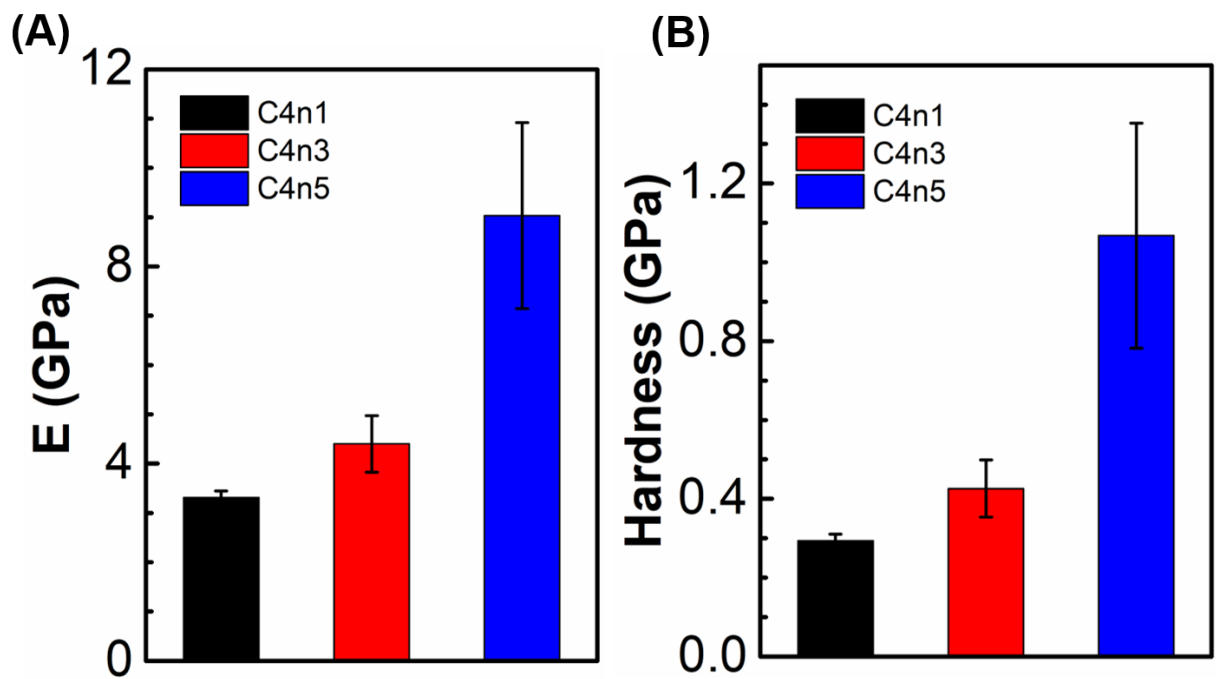

**Figure 4.** Out-of-plane Young's modulus (A) and hardness (B) as a function of n number.

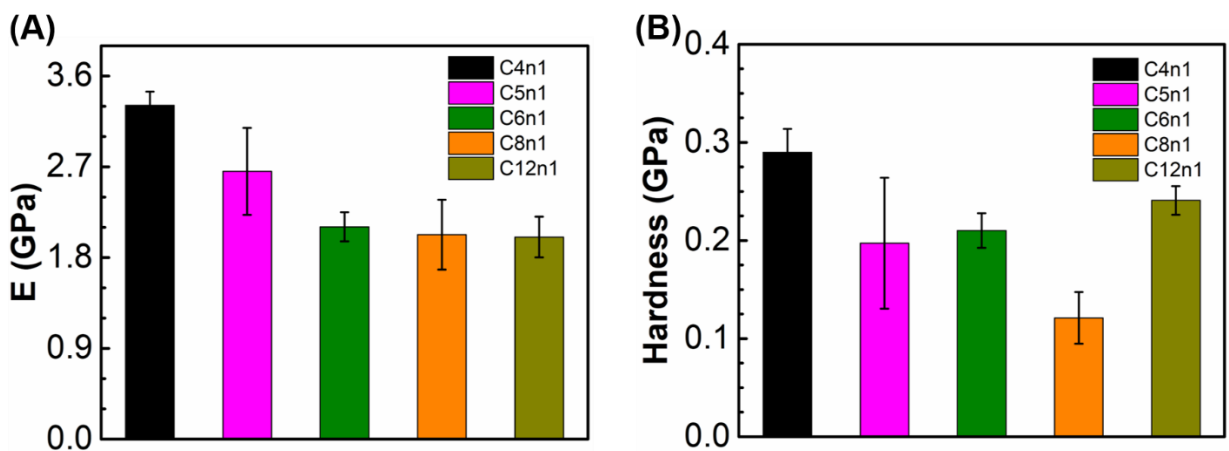

**Figure 5.** The effect of alkyl chain length on the out-of-plane Young's modulus (A) and hardness (B).

**Table I.** Experimental values of out-of-plane Young's modulus of 2D layered crystals

| 2D crystals | $E_\perp$ (GPa) | Methods | References |
|---|---|---|---|
| HOPG | 36.5 | ultrasonic and static tests | 53-54 |
| Muscovite Mica | 59 to 61 | Brillouin scattering and Nanoindentation | 55-57 |
| $MoS_2$ | 52 | Neutron and X-ray Scattering | 58 |
| h-BN | 24.5 to 27 | Brillouin scattering and X-ray Scattering | 59-60 |
| $Ti_3SiC_2$ | 320 | Nanoindentation | 61 |
| $Ti_3C_2T_x$ MXene (dry) | 12 to 70 | Contact Resonance AFM | 38 |
| $Ti_3C_2T_x$ MXene (wet) | 2 to 12 | Contact Resonance AFM | 38 |

**Table of Content**

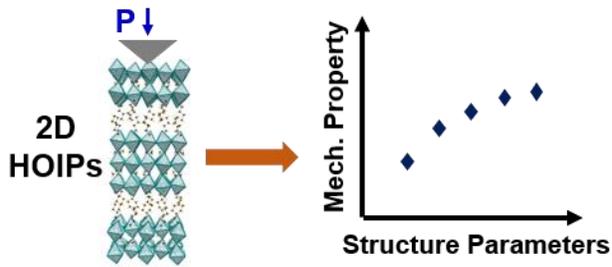